# Arguments for Detecting Octaneutrons in Cluster Decay of $^{252}$Cf Nuclei

G.N. Dudkin, A.A. Garapatskii and V.N. Padalko

Tomsk Polytechnic University, 634050 Tomsk, Russia

E-mail: dudkin@tpu.ru

**Abstract**

A new method of searching for neutron clusters (multineutrons) composed of neutrons bound by nuclear forces has been introduced and implemented. The method is based on the search for daughter nuclei that emerge at the nuclei cluster decay of $^{252}$Cf to neutron clusters. The effect of long-time build-up of daughter nuclei with a high atomic number and long half-life was utilized. As a result, the cluster decay of $^{252}$Cf to a daughter nucleus $^{232}$U (half- life of $T_{1/2}$= 68.9·years) was discovered. It is assumed that the emergence of $^{232}$U nuclei is attributed to emission of neutron clusters consisting of eight neutrons - octaneutrons. The emission probability of octaneutrons against α-decay probability of $^{252}$Cf is defined equal to $\lambda_C/\lambda_\alpha = 1.74 \cdot 10^{-6}$.

## 1. Introduction

The issue of stable nuclei (clusters) existence consisting of only neutrons has been the focus of experimental and theoretical research for 50 years. The detection of neutron nuclei would mean a discovery of a new type of nuclear matter bound with nuclear forces of $^x n$ neutrons, where $x$ is a number of bound neutrons. Such a discovery would have a great significance for all areas of nuclear physics as well as for astrophysics (neutron stars, *r*-process).

At present, the absence of bound nuclear stable states of the dineutron and trineutron is considered to be an established fact (see literature review [1] and references to it). The most number of experimental works was devoted to the issue related to the existence of the tetraneutron. There are several types of nuclear reactions, in which the tetraneutron search is possible. They are: exchange reactions; reactions of multinucleon transfers; fragmentation of target nuclei under the influence of light particles beams at high energy (some GeV); reactions of deep inelastic transfer; fragmentation of a projectile particle with a big excess of neutrons; and spontaneous and induced nuclear fission. A partial body of literature on methods and results of search for the tetraneutron is given in [2- 6]. The result of the tetraneutron search is negative.

As far as the majority of theoretical research focused on the tetraneutron existence is concerned, the results are also negative. The use of a standard *NN* potential does not allow to build a nuclear stable tetraneutron. The attempts to consider nonlocality or 3*N*-, 4*N*- interactions in calculations result to have neutrons bound in the tetraneutron, and properties of light nuclei in the stability valley change dramatically that contradicts the experimental data [7, 8]. However, the same authors admit the possibility of stable nuclei existence with a big excess of neutrons, particularly a magic nucleus $^8n$ - the octaneutron.

We have suggested a new method of search for neutron clusters. That is the use of cluster decay of $^{252}$Cf isotope. The cluster decay of heavy nuclei as a new type of natural radioactivity was discovered in 1984 [9].

In this process the masses of two fragments, as opposed to classical fission, differ greatly (6-12 times). This effect has been theoretically predicted in the paper [10], and the results of subsequent research have been summarized in the review [11].

Though there were some experimental difficulties in studying the cluster decay, it was still intensively investigated both experimentally and theoretically, and numerous interesting results have been achieved (see [11, 12] and references there). Here are some of the results.

1. All currently known nuclei exposed to the cluster decay refer to the range of heavy nuclei with the mass numbers of $A > 208$. The following types of cluster decays have been discovered: cluster $^{14}$C–parent nuclei: $^{221}$Fr, $^{221-226}$Ra, $^{225}$Ac; cluster $^{23}$F of $^{231}$Pa; clusters: $^{24-26}$Ne, –parent nuclei: $^{230,232}$Th, $^{231}$Pa, $^{232-234}$U; clusters $^{25,28,30}$Mg –: $^{234}$U, parent nuclei $^{237}$Np, $^{236,238}$Pu; clusters- $^{32,34}$Si –parent nuclei: $^{238}$Pu, $^{241}$Am.

2. The charge $Z$ and mass $A$ numbers of daughter nuclei produced by emission of clusters from heavy nuclei lie in narrow ranges: $80 < Z < 82$, $206 < A < 212$ (so called "lead radioactivity")

3. The kinetic energy of an outgoing particle is close to the so-called kinematic limit. It means that the particle takes away almost the whole decay energy. Consequently, after the decay the daughter nucleus is in either ground or exited state, but the excitation energy is not high.

4. The cluster decay probability is very low. The upper limit of $^{24,26}$Mg emission from $^{232}$Th nucleus is $T_{1/2} \geq 10^{29}$ s; the ratio of partial decay probabilities of $^{242}$Cm to $^{34}$Si and alpha particle is $\lambda_C/\lambda_\alpha = 10^{-15.8}$; a half-life of $^{14}$C emission from nucleus $^{223}$Ra $T_{1/2} = 10^{11}$ s and $\lambda_C/\lambda_\alpha = 10^{-8.9}$. The half-life of $T_{1/2}$ is connected with the decay constant $\lambda$ that characterizes the decay probability by relation $\lambda = \ln 2 / T_{1/2}$ s$^{-1}$.

For isotope $^{252}$Cf, theoretical estimates are provided by the following probability ratios for tracks: $^{252}$Cf→$^{14}$C+$^{238}$U, $T_{1/2} = 1.1 \cdot 10^{43}$ s, $\lambda_C/\lambda_\alpha = 10^{-35}$; $^{252}$Cf→$^{46}$Ar+$^{206}$Hg, $T_{1/2} = 1.1 \cdot 10^{32}$ s, $\lambda_C/\lambda_\alpha = \cdot 10^{-24}$ [13, 14]. Three experiments have been carried out searching for cluster decay of californium isotopes [15-17]. In experiment [15], for $^{249}$Cf→$^{50}$Ca+$^{199}$Pt reaction track, the following was received: $\lambda_C/\lambda_\alpha = 4.9 \cdot 10^{-9}$ and $T_{1/2} = 2.2 \cdot 10^{18}$ s. In experiment [16], just the upper limit for the cluster decay of $^{249}$Cf along tracks was set up: $^{249}$Cf→$^{44}$Ar+$^{205}$Hg; $^{249}$Cf→$^{46}$Ar+$^{206}$Hg; $^{249}$Cf→$^{48}$Ca+$^{201}$Pt; $^{249}$Cf→$^{50}$Ca+$^{199}$Pt - $\lambda_C/\lambda_\alpha \leq 1.5 \cdot 10^{-12}$ with the half-life of $T_{1/2} \geq 7.4 \cdot 10^{21}$ s. At the same time, theoretical calculations give ratios of partial relations from $\lambda_C/\lambda_\alpha = \cdot 10^{-20}$ to $\lambda_C/\lambda_\alpha = \cdot 10^{-28}$ [14]. In experiment [17], the upper limit for cluster emission of $^{46}$Ar or $^{48}$Ca in the cluster decay of $^{252}$Cf: $\lambda_C/\lambda_\alpha \leq \cdot 10^{-8}$ was set up. Emissions of light particles such as lithium, beryllium, and carbon isotopes were experimentally detected at the ternary fission of $^{252}$Cf nuclei.

However, there is no theoretical evidence of what emission probability for neutral clusters such as neutron clusters will be in case they exist.

**2. Experimental procedure**

In the case described here, to search for neutron clusters in the cluster decay of $^{252}$Cf, a neutron source on the basis of spontaneously fissionable isotope of $^{252}$Cf produced 37.5 years ago was used.

The search is based on the idea of utilizing the effect of long-time build-up of daughter nuclei with a high atomic number and a long half-life produced by the cluster decay of $^{252}$Cf. The search method includes measuring the spectrum of γ – quanta from the source and searching for γ –lines corresponding to daughter nuclei with a long half-life and a high atomic number.

The neutron source was placed inside a double-wall cylinder made of 1X18H10T stainless steel, with the total mass of 10.3 g, and hermetically sealed. The source mass of $^{252}$Cf at the moment of production was 6.27 micrograms. The initial number of nuclei of $^{252}$Cf is equal to $Nn=m·N_A/A$, where $m$ –source mass, $N_A$–Avogadro number, $A$– the number of nucleons in the nucleus (atomic number). In a certain period of time $t$, there were $N1(t)$ nuclei of $^{252}$Cf left in the source. The nuclei decrease was due to two processes: α-decay with the decay constant $\lambda_\alpha= 8.05289·10^{-9}$ s$^{-1}$ and spontaneous fission with the spontaneous fission constant $\lambda_{SF}=2.56849·10^{-10}$ s$^{-1}$. There were $N2(t)$ fissions for the same period of time. The calculation proves that at the start time there were $Nn=1.5·10^{16}$ nuclei of $^{252}$Cf. In 37.5 years $t=1.198368·10^9$ s, $N1=7.1·10^{11}$ nuclei remained in the source. Among the daughter nuclei produced during the 37.5 years of the cluster decay, only isotopes nuclei with a long half-life could have survived by the present time. The purpose of the experiment was to detect such nuclei.

For this purpose, the source was placed at a distance of 5.1 cm from the entrance gate of a semiconductor detector of γ – quanta made of high pure germanium HPGe (Canberra, model 1300), and the spectrum of γ – quanta was being measured for 96 hours (spectrometer loading ~8%).

The measuring chamber was surrounded by a 3-cm-thick layer of lead to neutralise the background load from environmental radioactive sources ($^{40}$K, $^{232}$Th, etc.). The effectiveness of γ – quanta registration was determined by $^{60}$Co, $^{152}$Eu, standard sources for γ – quanta energy intervals $E_\gamma$ = 119–2500 keV, and for the specified geometry is within the range of 0.65–0.09 %. The detector's energy resolution for $^{60}$Co lines ($E_\gamma$ =1173.24, 1332.5 keV) was determined as 1.8 keV. The background was also measured for 96 hours.

In the spectra 259 γ – lines of varying intensity were detected. The majority of the γ – lines belong to fission fragments with a short half-life. The search for long-living isotopes was carried out, isotope identification being made by γ – quanta energy and by the intensity ratio of the γ – quanta lines. The following isotopes were identified [18]: $^{102m}$Rh ($T_{1/2}$=3.742 years); $^{108m}$Ag ($T_{1/2}$=438 years); $^{125}$Sb ($T_{1/2}$=2.758 years); $^{133}$Ba ($T_{1/2}$=10.52 years); $^{134}$Cs ($T_{1/2}$=2.0652 years); $^{137}$Cs ($T_{1/2}$=30.08 years); $^{150}$Eu ($T_{1/2}$=36.9 years); $^{152}$Eu ($T_{1/2}$=13.537 years); $^{154}$Eu ($T_{1/2}$=8.593 years); $^{232}$U($T_{1/2}$= 68.9·years) $^{241}$Am ($T_{1/2}$=432.6 years); $^{243}$Am ($T_{1/2}$=7370 years); $^{243}$Cm ($T_{1/2}$=29.1 years); $^{249}$Cf ($T_{1/2}$=351 years); $^{251}$Cf ($T_{1/2}$=898 years).

The nuclei of $^{102m}$Rh, $^{108m}$Ag, $^{125}$Sb, $^{133}$Ba, $^{134}$Cs, $^{137}$Cs, $^{150}$Eu, $^{152}$Eu, $^{154}$Eu isotopes are apparently fission fragments. The presence of $^{241}$Am, $^{243}$Am, $^{249}$Cf, $^{251}$Cf isotopes in the source is due to the technology of $^{252}$Cf production in the high flux isotope reactor. The $^{232}$U isotope was detected in the spectrum for the γ – lines with energy $E_\gamma$ = 2614.5 keV and $E_\gamma$ = 583.2 keV owned isotope $^{208}$Tl, Fig 1.

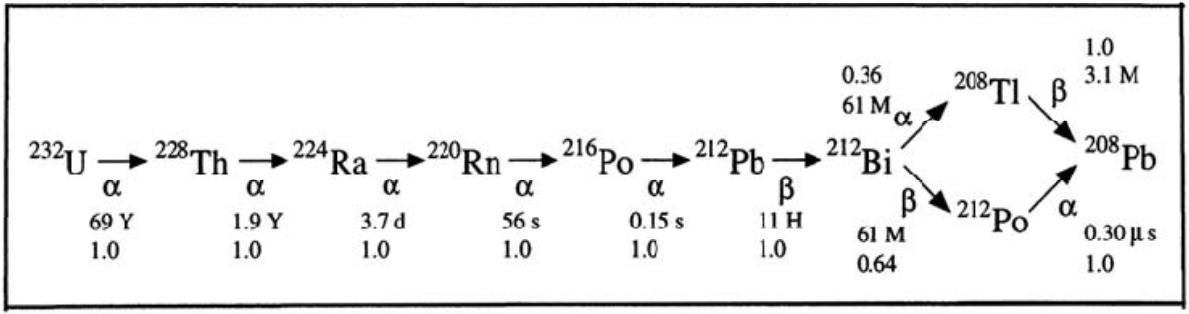

FIG 1. $^{232}$U decay chain. Decay channel: α, β. Half-life: Y-year, d- day, H- hour, M- minute, s- second. The probability of decay *P*: 1.0, 0.36, 0.64. $^{208}$Tl ($E_\gamma$=2614.5 keV, $I_\gamma$=99.75%; $E_\gamma$= 583.19 keV, $I_\gamma$ = 85.0%). $I_\gamma$ - quantum yield for the given isotope γ – line

Its presence is possibly due with the cluster decay: $^{252}$Cf→$^8n$+$^{244}$Cf($T_{1/2}$=19.4 M, $P_{\alpha.}$=100%)→$^{240}$Cm($T_{1/2}$= 27 d, $P_{\alpha.}$=99.5%)→$^{236}$Pu($T_{1/2}$= 2.86·Y, $P_{\alpha.}$=100%)→$^{232}$U

However, these lines are also present in the background spectrum, and their origin is related to thorium radioactivity ($^{232}$Th ($T_{1/2}$= 1.4·10$^{10}$·years, $P_{\alpha.}$=100%)→······→$^{208}$Tl ($T_{1/2}$= 3.1 min) →$^{208}$Pb.

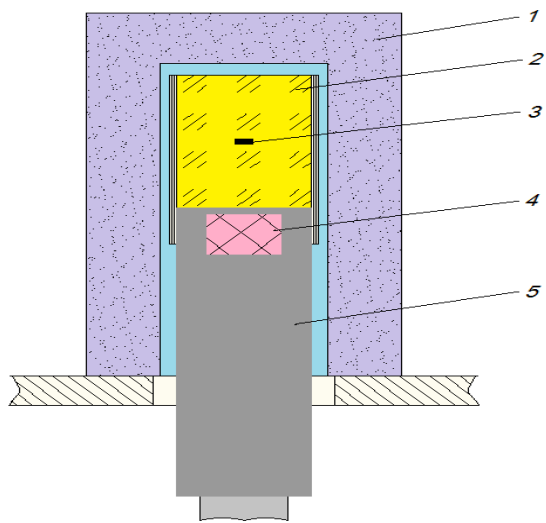

FIG. 2. 1- Lead shield; 2-Paraffin container; 3-Source of $^{252}$Cf; 4- Crystal HPGe; 5- Detector GC1020

Besides, as long as the $^{252}$Cf-based source has the neutron energy spectrum typical for the fission spectrum (i.e. the spectrum includes neutrons with MeV energies), reactions of inelastic scattering of neutrons are possible on isotopes of lead contained in a protective layer that surrounds the γ – quanta detector, including the direct excitation reaction – $^{208}$Pb(n,n`γ)$^{208}$Pb→ *(the most intensive lines: $E_\gamma$=2614.5 keV, $E_\gamma$=583.19 keV)*. To eliminate this effect, an extra experiment was carried out, Fig. 2.

The neutron source was placed into the geometrical centre of a cylinder-shaped container (diameter – 10 cm, height – 12 cm) and sealed with paraffin. This enabled to slow down fast neutrons to thermal energies, at which inelastic scattering reactions do not occur (reaction threshold equals 2.627 MeV [19]). Under these conditions, the energy spectrum of γ – quanta from the source was measured for 96 hours. After the subtraction of the background, the number of events in the spectrum under the $E_γ$=2614.5 keV and $E_γ$=583.5 keV γ – lines was processed in order to obtain information about the hypothetical probability of $^{252}$Cf cluster decay.

The number of nuclei of this isotope obtained from the experimental data was determined using the following formula:

$$N3_{exp} = \frac{S \cdot k_{YP} \cdot k_{St}}{\lambda_{CF} \cdot \varepsilon \cdot I_\gamma \cdot t_{meas}}, \quad (1)$$

where $S$ – the area under the γ – line peak in the spectrum (spectra were analyzed by means of the Genie-2000 software, Canberra); $\lambda_{CF}$ – nuclear decay constant; ε – detection efficiency for a γ – quantum with the corresponding energy; $I_\gamma$ –quantum yield for the given isotope γ – line[18]; $t_{meas}$ – the γ – spectrum measuring time with the germanium spectrometer; $k_P$, $k_{St}$ - calculated absorption coefficients of γ – quanta in paraffin and stainless steel of $^{252}$Cf source housing. The experiment results are presented in Table 1.

TABLE 1. Experiment results: $S_{NP}$, $S_{YP}$, $S_F$- the area full-energy peak in the spectrum, $A\gamma$ -γ – activity

| $E_γ$, keV $I_γ$ | No paraffin $S_{NP}$ | Yes paraffin $S_{YP}$ | Fon $S_F$ | Effect $S_{YP}$ - $S_F$ | ε | Absorption coefficient | | $A\gamma$, s$^{-1}$ | $N3_{exp}$ |
|---|---|---|---|---|---|---|---|---|---|
| | | | | | | $k_P$ | $k_{St}$ | | |
| 2614.5 35.9% | 6475±216 | 4375±114 | 3011±57 | 1364±127 | 9.5±0.5· 10$^{-4}$ | 1.24 | 1.1 | 15.8 ±1.6 | 4.95±0.52 ·10$^{10}$ |
| 583.5 30.6% | 5996±748 | 3686±668 | 1531±149 | 2155±684 | 3.9±0.2· 10$^{-3}$ | 1.49 | 1.6 | 12.3 ±4.1 | 3.85±1.27 ·10$^{10}$ |

For further study, the data for γ – quanta energy $E_γ$=2614.5 keV have been taken as more reliable.

Let us analyse californium decay for 3 channels:

1. Alpha-decay: $\lambda_\alpha$ = 8.05289·10$^{-9}$ s$^{-1}$ ($T_{1/2}$=2.729 years), probability $P0$=96.908%.
2. Fission: $\lambda_{SF}$ = 2.56849·10$^{-10}$ ($T_{1/2}$=85.574 years), probability $P1$=3.092%.
3. Hypothetical cluster decay channel: $^{252}$Cf ($N_1$, $\lambda_C$).

The solution of the equation set for the decay chains considering decrease in $^{252}$Cf through α-decay, fission and cluster decay produces formula (2) to calculate the number of accumulated daughter nuclei $N3$(t):

$$N3(t) = \frac{\lambda_C \cdot Nn \cdot (e^{-(\lambda_\alpha+\lambda_{SF}+\lambda_C)t} - e^{-\lambda_{CF} t})}{\lambda_{CF} - \lambda_\alpha - \lambda_{SF} - \lambda_C}, \quad (2)$$

Where, $\lambda_C$ – cluster decay constant; $\lambda_{CF}$ – decay constant for the daughter nucleus resulting from the cluster decay.

Having determined the ratio $\lambda_C/\lambda_\alpha$ = 10$^{-6}$, let us use formula (2) to calculate the number of accumulated daughter nuclei and their γ – activity $A\gamma$ depending on the daughter nuclei half-life, Fig. 3. The figure shows that the maximum probability for cluster decay detection by the γ – activity of daughter nuclei is reached for daughter nuclei with the half-life within the $T_{1/2}$ = 10÷100 years' range. $^{232}$U isotope, has a half-life of $T_{1/2}$= 68.9· years, which is optimal from the point of view of registration efficiency (Fig.3).

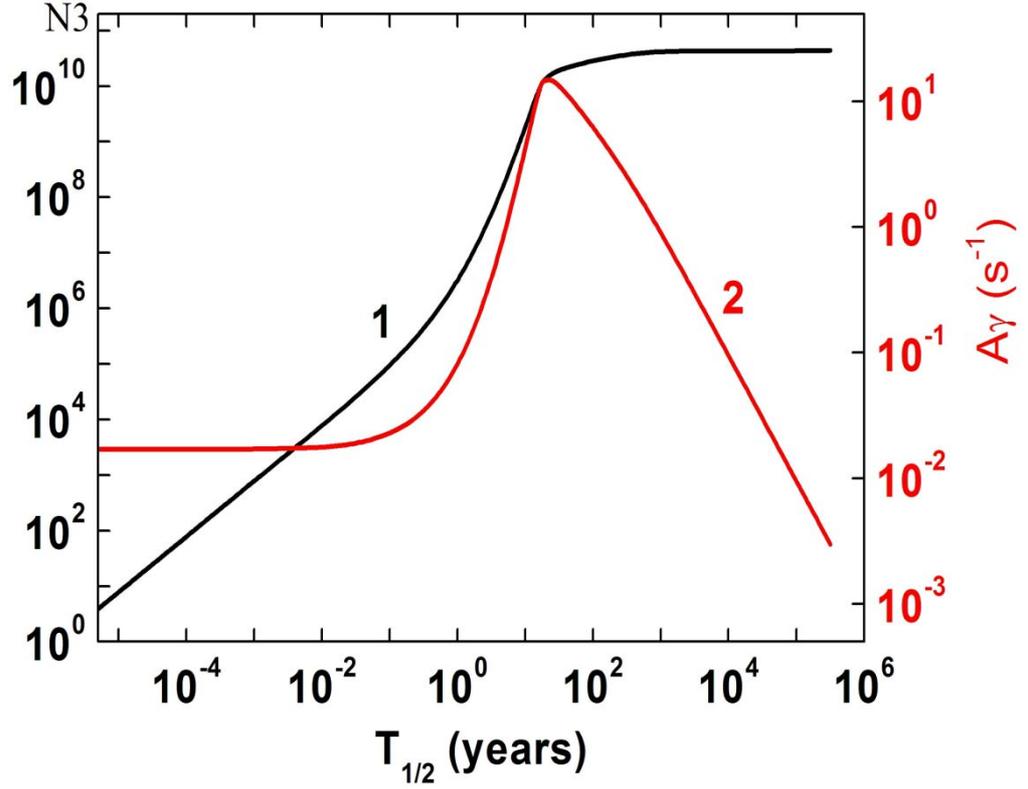

FIG. 3. 1 –number of accumulated daughter nuclei $N3$ (left scale) 2 –γ – daughter nuclei activity $A\gamma$, s$^{-1}$ (right scale), depending on a half-life $T_{1/2}$ of daughter nuclei.

## 3. Results and discussion

The fitting of formula (2), in which the only free parameter is $\lambda_C$ – cluster decay constant, to the experimentally obtained score of the nuclei number $N3_{exp}$, has given the following value $\lambda_C = 1.41\pm0.14\cdot10^{-14}$ s$^{-1}$. Hence $\lambda_C/\lambda_\alpha = 1.74\cdot10^{-6}$ and a half-life of $T_C = 1.61\pm0.16\cdot10^6$ years. Here statistical errors are given. They are determined by the number of events under the peak of γ- line $E_\gamma$=2614.5 keV, number of events in the "pedestal" under the peak, and number of background pulses in the same peak and then issued by spectra processing program (S501C, Genie-2000 software, Canberra). The error in registration efficiency ε, which is determined by activity accuracy of calibrating γ – sources from the manufacture, has also been considered.

What still are the possible nuclear reactions that can explain the origin of isotopes with γ – line $E_\gamma$=2614.5 keV?

1. The decay of $^{252}$Cf to α- particle. A long chain of radioactive transformations going with the α-decay of $^{252}$Cf and ending with $^{208}$Tl isotope passes the same chain as thorium radioactivity does - $^{232}$Th ($T_{1/2}$= 1.4·10$^{10}$·years, $P_\alpha$=100%)→······→$^{208}$Tl →$^{208}$Pb (stable) [18].

2. Binary fission of $^{252}$Cf and formation of radioactive fission fragments emitting γ- quanta with energy close to $E_\gamma$=2614.5 keV. These include [18]: $^{74}$Br ($E_\gamma$=2615.2 keV, $I_\gamma$=7.37%), $^{93}$Rb ($E_\gamma$=2614.1 keV, $I_\gamma$=0.149%), $^{104}$Ag ($E_\gamma$=2613.4 keV, $I_\gamma$=0.055%), $^{135}$Te ($E_\gamma$=2615.5 keV,

$I_\gamma$=0.06%), $^{120}$I ($E_\gamma$=2613.0 keV, $I_\gamma$=0.25%), $^{139}$Xe ($E_\gamma$=2613.7 keV, $I_\gamma$=0.034%), $^{141}$Cs ($E_\gamma$=2615.5 keV, $I_\gamma$=0.114%), $^{136}$Pr ($E_\gamma$=2613.2 keV, $I_\gamma$=0.105%), $^{148}$Tb ($E_\gamma$=2614.3 keV, $I_\gamma$=0.69%).

3. Cluster decay: either $^{252}$Cf→ $^{16}$Be + $^{236}$Pu ($T_{1/2}$= 2.86· years, $P_\alpha$=100%)→$^{232}$U ($T_{1/2}$= 68.9·лет, $P_\alpha$=100%)→…..$^{208}$Tl →$^{208}$Pb (stable); or $^{252}$Cf→ $^{20}$C +$^{232}$U($T_{1/2}$= 68.9· years, $P_\alpha$=100%)→……$^{208}$Tl →$^{208}$Pb (stable).

4. Cluster decay $^{252}$Cf→$^{44}$P +$^{208}$Bi (T$_{1/2}$ =3.68·10$^5$ y, $E_\gamma$=2614.5 keV, $I_\gamma$=99.78%).

5. The process of the following type: $^{252}$Cf→8n + $^{244}$Cf($T_{1/2}$=19.4 minutes, $P_\alpha$=100%)→$^{240}$Cm($T_{1/2}$= 27 days, $P_\alpha$=99.5%)→$^{236}$Pu($T_{1/2}$= 2.86· years, $P_\alpha$=100%)→$^{232}$U($T_{1/2}$= 68.9·years, $P_\alpha$=100%)→…. Then follows the same chain of decays as in case 3. That is, in the process of fission, the nucleus of $^{252}$Cf "slowly moves" to a breaking point and this time is enough to radiate an excess of excitation energy in the form of eight prescission neutrons! As a result of evaporation of eight neutrons, the fission process stops, and a weakly excited nucleus of $^{244}$Cf is formed. What is the probability of such a process?

6. The impact of impulses summation effect owing to an accidental coincidence of γ- quanta in time from various radioactive isotopes as well as summation peaks owing to cascade γ- quanta. Such effects have not been found for any isotopes detected in the spectrum.

The final results of background processes are shown in Table 2.

TABLE 2. Contribution of all possible background processes to γ – line intensity with the energy of $E_\gamma$=2614.5 keV. Only those fission fragments have been considered for which nonzero probability of $P$ cumulative output exists [20].

| Reaction | $T_{1/2}$ | λ, s$^{-1}$ | $E_\gamma$, keV | $I_\gamma$, % | P | N3, N2 | Aγ, s$^{-1}$ |
|---|---|---|---|---|---|---|---|
| $^{252}$Cf→$^8$n(8n)+…$^{232}$U…….. | 68.9 y | 3.19·10$^{-10}$ | 2614.5 | 36 | - | 4.95±0.51 ·10$^{10}$ | 15.8±1.6 |
| $^{252}$Cf→α+…..…$^{232}$Th……. | 1.405·10$^{10}$ y | 1.6·10$^{-18}$ | 2614.5 | 36 | - | 1.45·10$^{16}$ | 0.023 |
| $^{252}$Cf→$^{208}$Bi +$^{44}$P | 3.68·10$^5$ y | 5.97·10$^{-14}$ | 2614.5 | 99.78 | - | 2.5·10$^{10}$ | 0.0015 |
| $^{252}$Cf→$^{93}$Rb+ | 5.9 s | 0.117 | 2614.1 | 0.149 | 4.7·10$^{-3}$ | 7 | 0.0013 |
| $^{252}$Cf→$^{135}$Te+ | 19 s | 3.65·10$^{-2}$ | 2615.5 | 0.061 | 1.9·10$^{-2}$ | 95 | 0.002 |
| $^{252}$Cf→$^{139}$Xe+ | 39.7 s | 1.74·10$^{-2}$ | 2613.7 | 0.034 | 3.9·10$^{-2}$ | 406 | 0.0024 |
| $^{252}$Cf→$^{141}$Cs+ | 24.9 s | 2.78·10$^{-2}$ | 2615.5 | 0.114 | 4.8·10$^{-2}$ | 314 | 0.0096 |

The calculation of the fission fragments contribution to γ – line intensity with the energy of $E_\gamma$=2614.5 keV was made in accordance with the following formula:

$$N2(t) = \frac{\lambda_{SF} \cdot Nn \cdot P \cdot (e^{-(\lambda_\alpha + \lambda_{SF})\cdot t} - e^{-\lambda_{FF}\cdot t})}{\lambda_{FF} - \lambda_\alpha - \lambda_{SF}} \qquad (3)$$

Here: $\lambda_{FF}$ - fission fragment decay constant; $P$ – probability of cumulative nuclei formation of the isotope under fission.

Possible channels of a cluster decay (or prescission neutrons evaporation) from $x$=2 to $x$=12 have also been considered. For low cluster multiplicity, the decay of $^{252}$Cf to clusters $x$ = 2.3.4 is not studied since $^{250}$Cf, $^{249}$Cf, $^{248}$Cf nuclei are involved into the process of $^{252}$Cf accumulation in a high-flux reactor, and therefore their "tracks" have been saved in the source made. For high cluster multiplicity, the decay of $^{252}$Cf to clusters $x\geq$ 12 is not studied since the database of Nuclear Data Services of IAEA does not contain reliable data about properties of $^{\leq 240}$Cf nuclei. In the majority of channels (except the channel x= 8), the detection of the effect of neutron clusters emission failed: either due to the overlapping of desired γ- lines with γ- lines of isotopes nuclei that appear in the source as a result of other processes, or due to the absence of analytical lines of γ- quanta with a high quantum yield.

Probabilities of processes 3, 5, are defined by one and the same value $\lambda_C = 1.41 \pm 0.07 \cdot 10^{-14}$ s$^{-1}$ obtained experimentally. The question arises – which of the processes should one give preference to? It is clear that such a decay constant does not fit processes:

$$^{252}\text{Cf} \to {}^{16}\text{Be} + {}^{236}\text{Pu} \to {}^{232}\text{U} \to \ldots \to {}^{208}\text{Tl} \to {}^{208}\text{Pb (stable)} \quad (4)$$

$$^{252}\text{Cf} \to {}^{20}\text{C} + {}^{232}\text{U} \to \ldots \to {}^{208}\text{Tl} \to {}^{208}\text{Pb (stable)} \quad (5)$$

Since, it contradicts all the set of currently obtained experimental and theoretical data for $^{252}$Cf cluster decay probabilities. There is no any information about the probability of the emission process for eight prescission neutrons and fission process termination.

$$^{252}\text{Cf} \to 8n + {}^{244}\text{Cf} \to \ldots \to {}^{232}\text{U} \to \ldots \to {}^{208}\text{Tl} \to {}^{208}\text{Pb (stable)} \quad (6)$$

It is only known that the portion of prescission neutrons is not more than 10-15% of all neutrons emitted in the process of one fission event [21, 22]. The probability of eight neutrons emission for one fission event is $P_n = 4.5 \cdot 10^{-3}$ [23].

We suppose that the most probable process is the neutron cluster emission consisting of eight neutrons – octaneutron.

$$^{252}\text{Cf} \to {}^{8}n + {}^{244}\text{Cf} \to \ldots \to {}^{232}\text{U} \to \ldots \to {}^{208}\text{Tl} \to {}^{208}\text{Pb (stable)} \quad (7)$$

This conclusion does not contradict the results of work [24], during which the experimental evidence of neutron clusters formation with multiplicity $x \geq 10$ and $x \geq 11$ in induced fission of $^{235}$U in a nuclear reactor was found. The work is aimed at attracting physicists' interest to the issue of stable neutron matter existence.

**Acknowledgments**

This work was supported by the Ministry of Education and Science of the Russian Federation, Project № 2.1704.2011.

We would like to express our deep gratitude to our laboratory colleagues V.M. Golovkov and

V.A. Varlachev for their assistance in doing this research and providing useful advice as well as critiques. We are also sincerely and heartily grateful to Tamara G. Petrashova, our language advisor for her guidance in the linguistics of English.